\documentclass{PoS}

\title{FLUX AND SPECTRAL VARIATIONS OF 1E{\thinspace}1740.7$-$2942 OVER THE YEARS 2003$-$2012}

\ShortTitle{1E{\thinspace}1740.7$-$2942 with INTEGRAL}

\author{
        \speaker{Manuel Castro}, 
                 Flavio D'Amico, 
                 Francisco Jablonski, 
                 Jo\~ao Braga                                                                                    \\
                 Instituto Nacional de Pesquisas Espaciais (INPE), S\~ao Jos\'e dos Campos, S\~ao Paulo, Brazil  \\
                 E-mail: \email{manuel.castro@inpe.br}
       }
\newcommand{\E}{1E{\thinspace}1740.7--2942}                  
\abstract{ 
          The black hole system \E\ is usually the brightest hard X-ray 
          source (above 20 keV) near the Galactic Center, but presents
          some epochs of low emission (below the INTEGRAL detection 
          limit, for example). In this work, we present the results of studies on 
          \E\ over 10 years, using the instruments ISGRI/IBIS and 
          JEM-X, both on board the INTEGRAL observatory. We fit the spectra with both
          the \textsf{compTT} and \textsf{cutoffpl} models. According to the fits and taking 
          the mean value over the 10 years, we have obtained a plasma 
          temperature in the range $\sim$20--90\thinspace keV, 
          and an average powerlaw index of 1.41 ($\sigma$=0.25).
          We have also made a Lomb--Scargle periodogram of the flux in the
          50--20{\thinspace}keV band and found two tentative periods at 2.90 and
          3.99{\thinspace}days. 
          We present here the preliminary results of this ongoing work.
         }

\FullConference{10th INTEGRAL Workshop: "A Synergistic View of the High Energy Sky" - Integral2014,\\
		15-19 September 2014\\
		Annapolis, MD, USA }

\begin{document}
%
%
%
\section{Introduction}

The source {\E} is a black hole candidate nearby the center of our Galaxy. It was
discovered by the Einstein satellite \cite{Hertz1984} and it was 
thoroughly studied by several high energy missions like SIGMA/GRANAT, INTEGRAL, Suzaku and Chandra \cite{Syunyaev1991,Bouchet2009,Reynolds2010,Gallo2002}.
 It was also the subject of a recent study by our group \cite{Castro2014}, where we modelled the soft--to--hard X-ray spectrum
as due to (thermal) Comptonization. 

The INTEGRAL satellite \cite{Winkler2003} has already provided an uniform observation database of
\E. This database is suitable for a study of any possible long--term variability of the source. One interesting
goal of such a study is the search for modulations in the X-ray lightcurve to reveal any
possible orbital period. Reported orbital periods for {\E} \cite{Smith1997,Smith2002} are still
tentative, and a more precise determination of this period would be very important in the search for
a (longer wavelength) counterpart of {\E}, a challenging task that remains to be accomplished. 

In this work we show the results of an ongoing effort to determine the orbital period
of {\E} (if there is one). We have made use of the INTEGRAL database consisting of all of the {\E} observations from 2003
up to 2012.  We have found two marginally significant tentative periods
at ${\sim}${\thinspace}2 and ${\sim}${\thinspace}4 days. In the next sections we describe
our database, our data reduction and analysis, and discuss our results. 

\section{Data Reduction and Analysis}

\begin{figure}[h!]
\begin{center}
\includegraphics[angle=0,width=4cm]{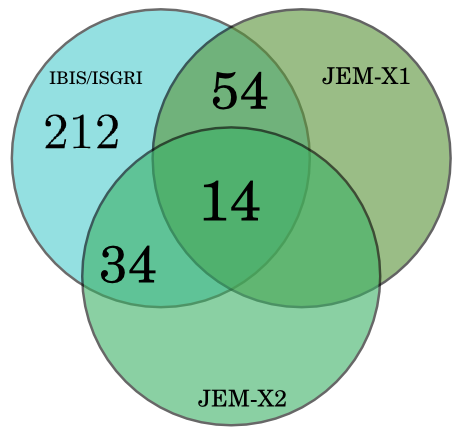}
\end{center}
\caption{
                  Our database of 314 retrieved observations of {\E}.
                  In the vast majority of cases (212) the source was imaged by IBIS alone. 
                  In 14 occasions the source was imaged by the 2 JEM-X units plus IBIS. In 54 with
                  IBIS plus only the JEM-X1 unit and 34 with IBIS plus JEM-X2 unit. 
                 }
\label{fig01}
\end{figure}

From the public archive of INTEGRAL, we downloaded all of {\E} observations starting at 2003
up to 2012. {\E} spends most of its time in the so-called canonical low--hard state (LHS), where
the fraction of the total flux is higher in the hard X-ray band than in the soft X-ray band. Accordingly, the source flux is
below the sensitivity of JEM-X detectors in their energy range, being preferably monitored in the band
covered by the IBIS telescope. As a result, Figure{\thinspace}({\ref{fig01}}) shows how our
retrieved 314 observations database is distributed between the 3 INTEGRAL imaging (unit) telescopes. 

Spectra were extracted for each individual revolution using an automatic extraction script
based on the OSA 10.0 software. In some cases the spectra are built only with IBIS data, covering
the 20--200 keV energy band. When the source was also imaged by
JEM-X, a spectrum from 5--200 keV could be extracted with the
aid of our automatic script. After spectrum extraction, data were fitted also automatically 
with the aid of a Tcl script. For this preliminary study, we fitted the 20--200 or the 5--200{\thinspace}keV
spectra (when present) with two models. The first one is a simple 
{\sf cutoffPL}, suitable, for instance, to determine in which
state the source was observed \cite{Remillard2006}. The second model is 
{\sf compTT}, following the procedures of our previous study of the source
\cite{Castro2014}.  In 14 occasions, as is shown in
Figure ({\ref{fig01}}), the spectrum extraction script needs to run three times
(first run is always for IBIS), since both JEM-X units
provided data. After fitting the data, our Tcl script saves the fit parameters
(with error estimates) and calculated fluxes.  We built our database with this kind of datafiles.
In this first stage of our study, for simplicity, errors in the flux measurements are assumed to be
within ${\pm}${\thinspace}5{\%}.
\section{Results}

\begin{figure}[h!]
\includegraphics[angle=270,width=15cm]{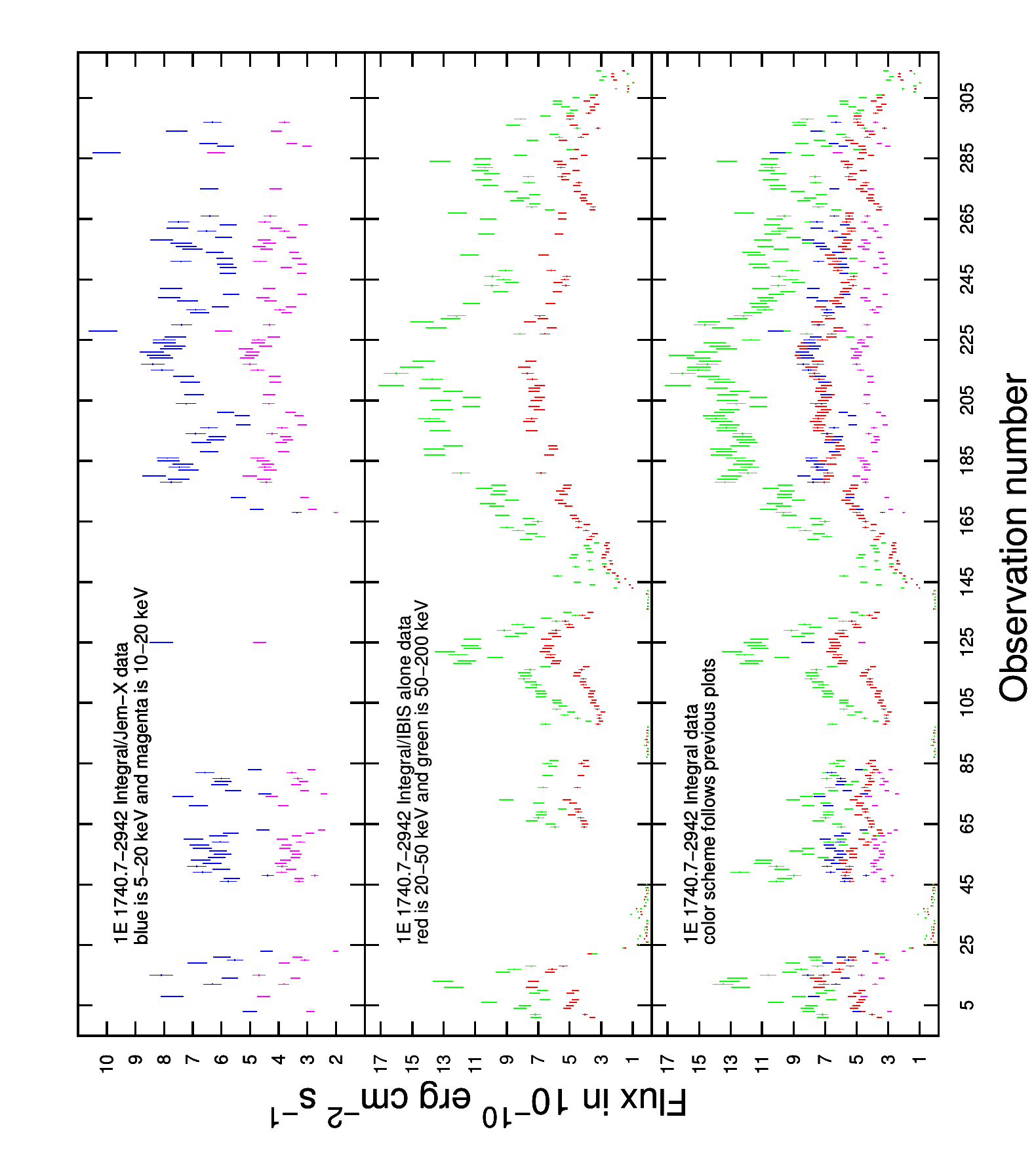}
         \caption{
                  The flux history of {\E} based on our database.
                 }
\label{fig03}
\end{figure}

\begin{figure}[h!]
\begin{center}
 \includegraphics[angle=0,scale=0.5]{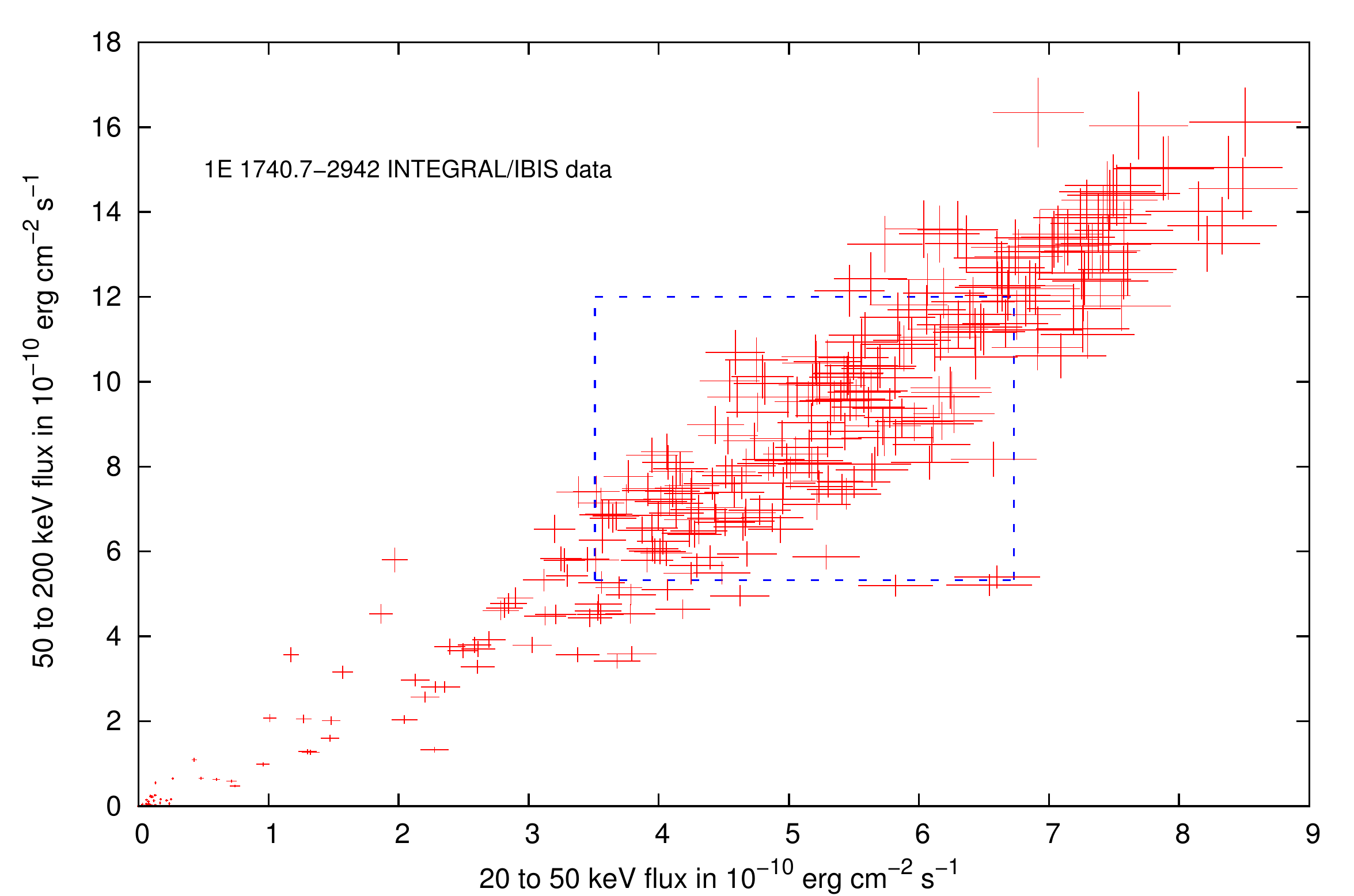}
\end{center}
         \caption{
                  The fluxes in the two highest energy bands in this study. Shown in the
                  plot also (in blue) is the region of average{\thinspace}$\pm${\thinspace}1${\sigma}$
                  for both bands. This shows the region (in flux) of preferred fluxes for the 
                  (canonical) high state where {\E} is commonly observed. 
                 }
\label{fig04}
\end{figure}

\begin{figure}[h!]
\begin{center}
\includegraphics[angle=0,scale=0.6]{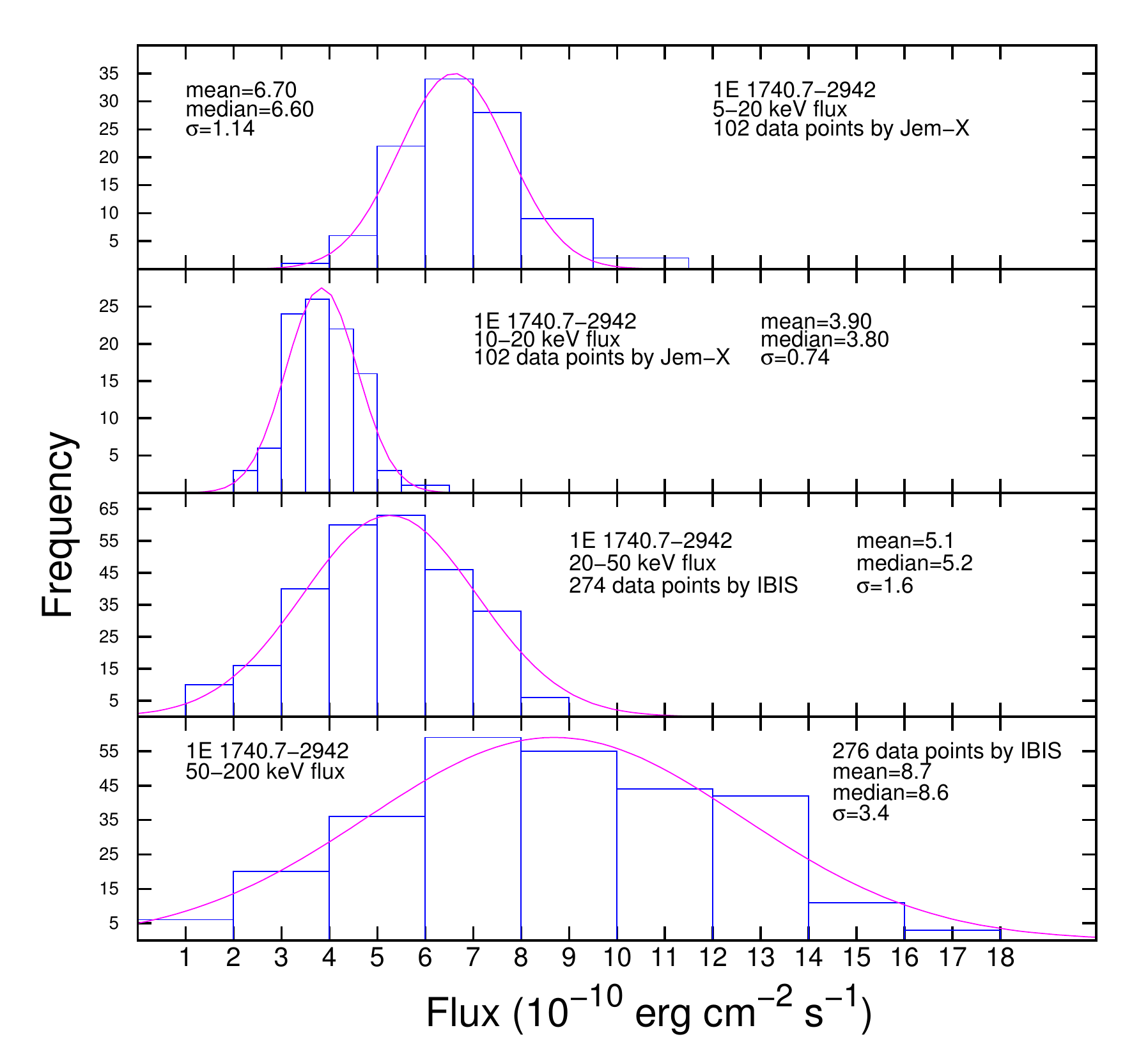}
\end{center}
         \caption{
                  Histogram distribution for fluxes in the 4 energy bands considered in this study.
                 }

\label{fig05}
\end{figure}

\begin{figure}[h!]
\includegraphics[angle=0,scale=0.6]{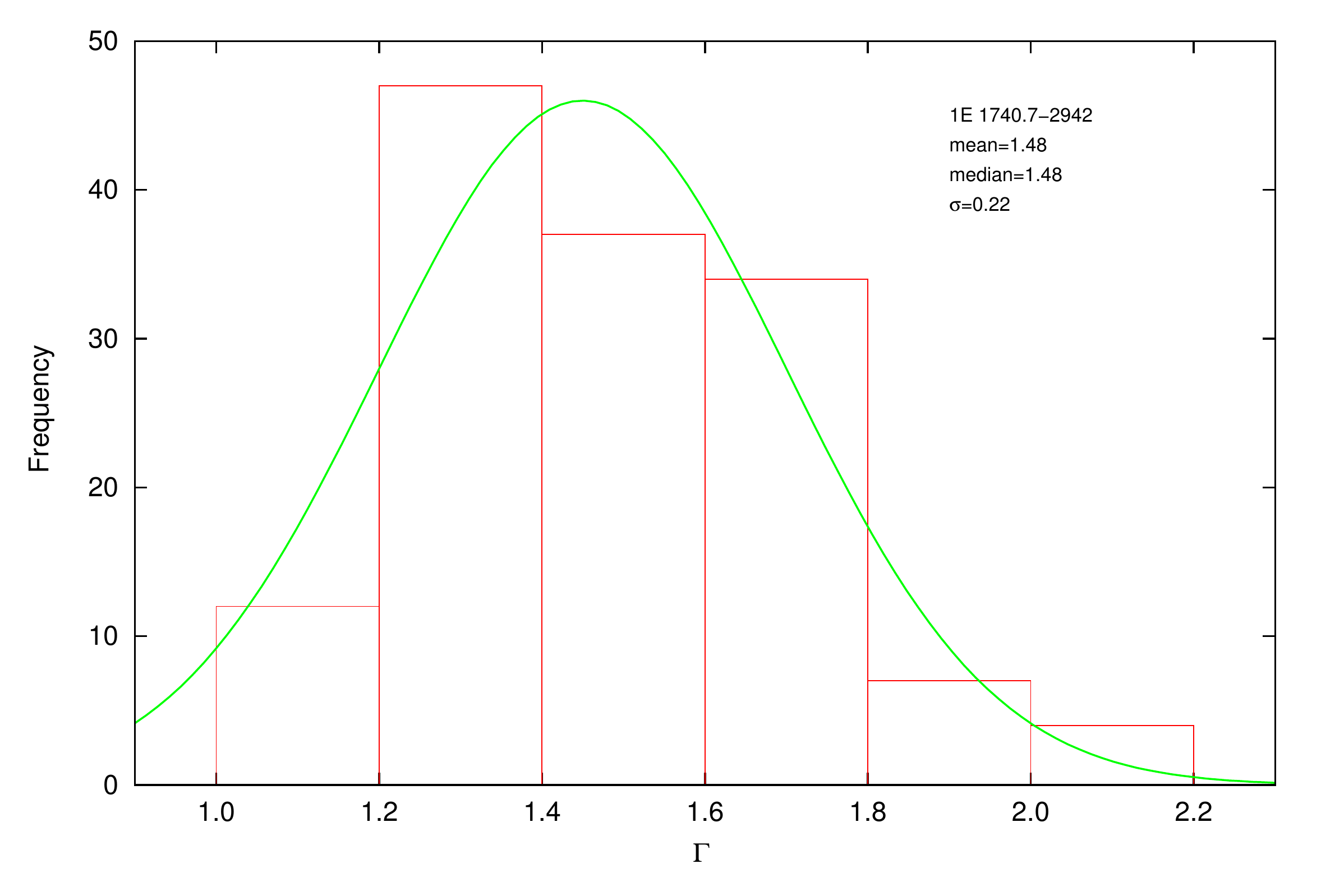}
         \caption{
                  Histogram distribution for the powerlaw index ${\Gamma}$ derived from a 
                  {\sf cutoffPL} fit in XSPEC.
                 }
\label{fig06}
\end{figure}

\begin{figure}[h!]
\includegraphics[angle=0,scale=0.6]{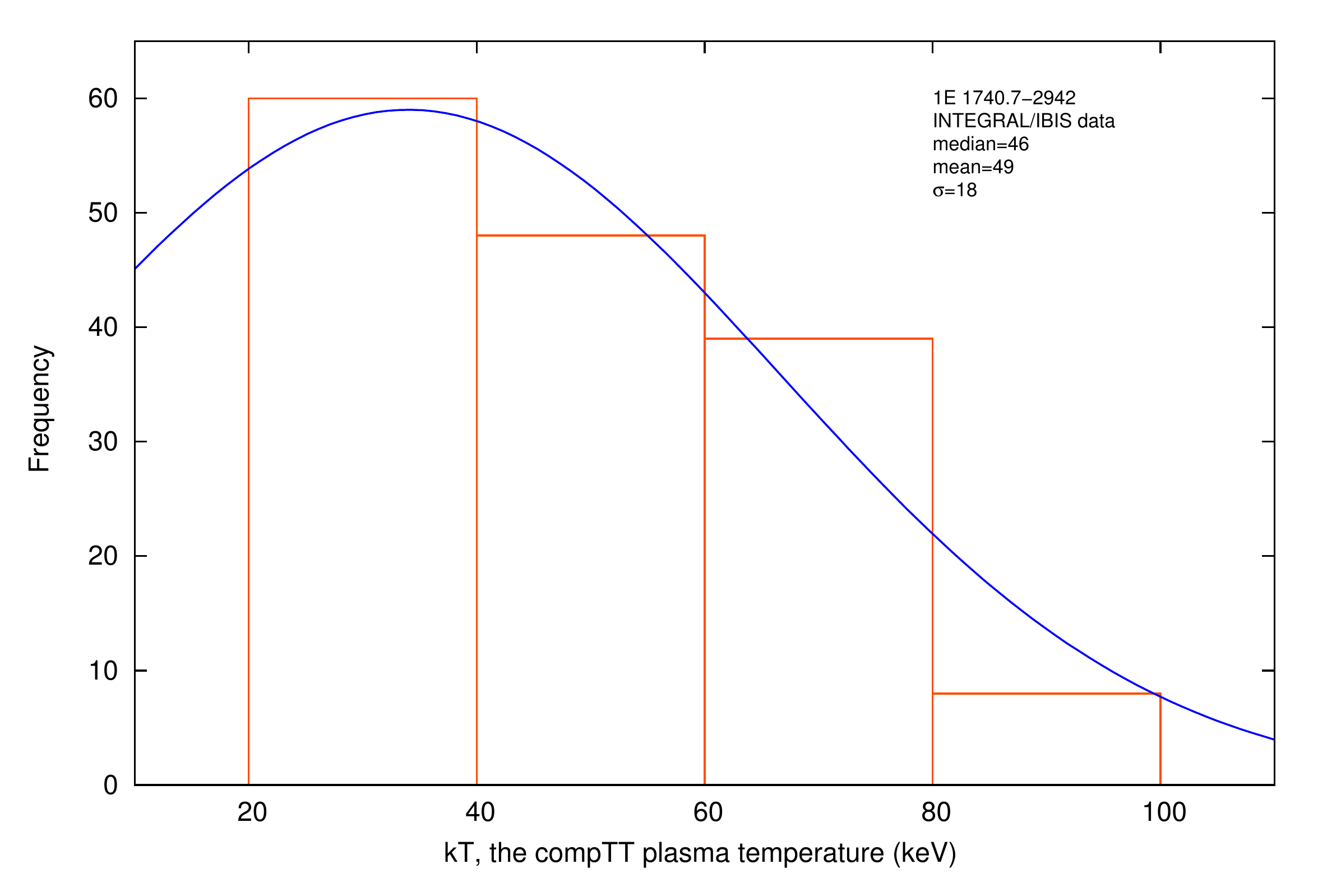}
         \caption{
                  Histogram distribution for the plasma temperature kT derived from a 
                  {\sf compTT} fit in XSPEC.
                 }
\label{fig07}
\end{figure}

\begin{figure}[h!]
\includegraphics[angle=0,scale=0.6]{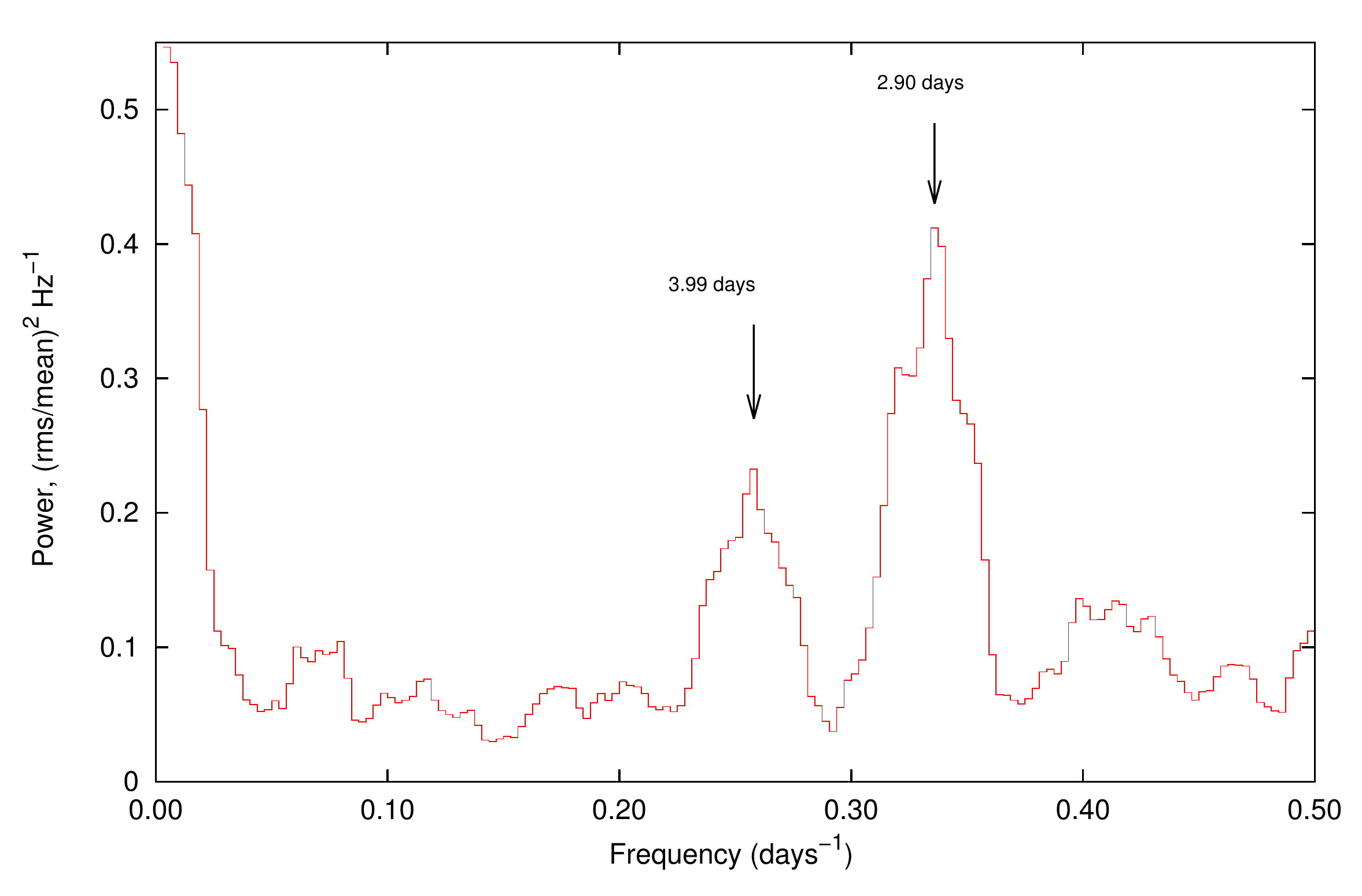}
         \caption{
                  Lomb-Scargle periodogram of the 50--200{\thinspace}keV flux showing two tentative
                  possible periods for {\E} (see text for details).
                 }
\label{fig08}
\end{figure}

The {\E} flux history of our database is shown in Figure{\thinspace}({\ref{fig03}}).
In Figure{\thinspace}({\ref{fig04}}) we show the correlation between the fluxes in the two highest energy
bands of this study. The histogram distributions for the 4 bands we
adopted in this study are presented in Figure{\thinspace}({\ref{fig05}}). The powerlaw index
and plasma temperature data for our database are presented in 
Figures{\thinspace}({\ref{fig06}}){\thinspace}and{\thinspace}({\ref{fig07}}), respectively.\\
We have also produced a Lomb-Scargle periodogram using the R software environment \cite{R}. This is shown in Figure{\thinspace}({\ref{fig08}}).
\section{Discussion}

Our results have confirmed, as it is widely known, that {\E} spends most of its time in its canonical LHS, as derived from
the powerlaw index of a fit with {\sf cutoffPL} (see Figure {\ref{fig06}}). Spectral fitting in our database from 20 to 200 keV or 5 to 200{\thinspace}keV (when available) using
{\sf compTT} models, yields an average temperature of the electronic plasma, which (thermally) Comptonizes 
the soft seed photons, of ${\sim}${\thinspace}50{\thinspace}keV (see Figure{\thinspace}{\ref{fig07}}). 
The ${\sim}${\thinspace}300 points in our database were helpful to better constrain this value. It must be emphasized, however,
that at this point of our preliminary
study the data for deriving ${\Gamma}$ and kT$_{\hbox{\scriptsize{e}}}$ are based on very low fluxes (see Figure{\thinspace}{\ref{fig03}}) so the uncertainties are somewhat high.

As mentioned above,
one of the main goals in this work is to search for possible binary periods. Our database is one
of the best for such purposes, since {\E} spends most of its time in the LHS,
i.e., the flux of {\E} is higher in the hard X-ray band ($E${\thinspace}$>$20{\thinspace}keV). Previous searches
up to date where limited by databases in softer bands. INTEGRAL and the telescopes we used
in this study, JEM-X and IBIS, are also imaging instruments and consequently {\E} flux is not affected by
source confusion. Previous studies of searches for {\E} periods have made use (mostly) of data collected
by non-imaging instruments and at the soft X-ray band ($E${\thinspace}$<$20{\thinspace}keV). 

Our
results, which can be seen in (Figure{\thinspace}{\ref{fig08}}), show the presence of periods at approximately 2.90 and 3.99{\thinspace}days. Even though relatively well above local noise level in the periodogram, the second period must be taken with caution since it is quite
close to a multiple of the one-day sampling period used. In addition, the 2.90-day period produces a folded light
curve that is not very clean, which introduces uncertainties that we are still analysing. It must also be emphasized that much more sensitive
studies using RXTE \cite{Smith2002} have shown a period of $12.73${\thinspace}${\pm}${\thinspace}days.

\section{Conclusion}

We have analysed data from the INTEGRAL mission to build flux and spectral history of {\E}
from 2003 to 2012. Our data analysis is automatic
and divided in two main fronts. First we use a script to automatically
reduce the data using tasks of the OSA 10.0 software. Second, a Tcl script fits
the data with XSPEC--aided commands.  We collected, when possible, flux
in 4 energy bands. Our data show, based on our fits with 
{\sf cutoffPL}, that the source is observed to be in its canonical LHS state, where the
flux in the hard X-ray band is greater than in the soft one. Our characterization
of the {\E} spectra with {\sf compTT} shows an average kT$_{\hbox{\scriptsize{e}}}$ of 50{\thinspace}keV. A Lomb--Scargle periodogram analysis has shown two tentative periods of 2.90 and 3.99{\thinspace}day. These results must be taken with caution due to uncertainties that are still being addressed.

\end{document}